\documentclass[twocolumn,showpacs,preprintnumbers,amsmath,amssymb]{revtex4}
\usepackage{graphicx}
\usepackage{bm}

\newcommand{\ie}{{\it i.e.}}
\newcommand{\eg}{{\it e.g.}}

\def\ket#1{\mid\!#1\rangle}

\def\ave#1{\langle~#1~\rangle}

\begin{document}

\title{Dynamics for partially coherent Bose-Einstein condensates in double wells}

\author{Li-Hua Lu  and You-Quan Li}
\affiliation{Department of Physics and Zhejiang Institute of Modern
Physics, Zhejiang University, Hangzhou 310027, People's Republic of
China}


\begin{abstract}
The dynamical properties of partially coherent Bose-Einstein
condensates in double wells are investigated in three typical
regimes.
In the extreme Fock regime, the time evolution of the
degree of coherence is shown to decay rapidly.
In the Rabi regime,
a relation between the amplitude of Rabi oscillation and
the degree of coherence is obtained, which is expected to determine the degree
of coherence by measuring the amplitude of Rabi oscillation.
The study on the self-trapping phenomena in the Josephson regime
exhibits
that both the degree of coherence and the initial relative phase can
affect the final particle distribution.

\end{abstract}
\received{\today} 


\maketitle
\section{Introduction}

Quantum tunneling through a barrier, as a paradigm of quantum
mechanics, has been observed in different systems, such as two
superconductors separated by a thin insulator~\cite{Likharev}, two
reservoirs of superfluid helium connected by nanoscopic
apertures~\cite{Pereverzev,Sukhatme}, and Bose-Einstein condensates
(BECs) trapped in double wells~\cite{Albiez}. Comparing with the
former two systems, BECs in double wells offer a versatile tool to
study the quantum tunneling phenomena due to the fact that almost
each parameter, such as the inter-well tunneling strength, the
inter-particle interaction, and the energy bias between the two
wells, can be tuned experimentally. The competition between those
parameters makes BECs in double wells exhibit several fascinating
phenomena, like Rabi oscillation, self-trapping, and Josephson
oscillation, which have been extensively
studied~\cite{smerzi,Milburn,Raghavan,Jai}. When a condensate is
expected to be employed as a qubit, the first obstacle attempted to
be avoided is the decoherence which was studied in experiment
through the interference between BECs~\cite{T.Schumm,S.Hoffer}.
Theoretically, the effect of decoherence on the dynamics of BECs in
double wells was recently discussed with the help of single-particle
density matrix~\cite{Wang}. The problem of decoherence in qubit
measurement was recently investigated~\cite{Soko} with a BEC in
double wells. In current literature, the most of authors paid more
attention on the time evolution of the atom distribution but less
attention on the relative phase between the condensates in two
wells. Meanwhile, the systems were mostly assumed to be completely
coherent at the initial time. Since both relative phase and degree
of coherence are believed to affect the dynamical properties of
BECs in double wells, it is worthwhile to study the coherence
dynamics with attention to the relative phase and the degree of
coherence.

In this paper, we study a BEC system in double wells with different
degrees of coherence. In mean-field approximation, we study the
dynamical properties of the system in different regimes with the
help of  the single-particle density matrix and show that, in
comparison to the completely coherent case, partially coherent BECs
in double wells can exhibit richer physics. In the next section, we
model the partially coherent system and give the mean-field
dynamical equation for the elements of single-particle density
matrix. In Sec.~\ref{sec:coherence}, we study the evolution of the
degree of coherence and discuss the dynamical property of the
partially coherent system in Fock regime. In Sec.~\ref{sec:rabi}, we
study the system in Rabi regime and discuss the influence of degree
of coherence on the Rabi oscillation. In Sec.~\ref{sec:selftrap}, we
study self-trapping phenomenon for the partially coherent system.
Then brief summary and discussion are given in Sec.~\ref{sec:sum}.

\section{Modeling partially coherent systems}\label{sec:model}

We consider a Bose-Einstein condensate confined in a double-well
potential, where atoms can tunnel between the two wells.
The Hamiltonian of such a system in the second-quantization form is given by,
\begin{eqnarray}\label{eq:hamquan}
\hat{H}=\frac{\gamma}{2}(\hat{n}^{}_1-\hat{n}^{}_2)
-T(\hat{a}_1^\dagger\hat{a}^{}_2+\hat{a}_2^\dagger\hat{a}^{}_1)+\frac{U}{4}
(\hat{n}^{}_1-\hat{n}^{}_2)^2,\nonumber\\
\end{eqnarray}
where the bosonic operators $\hat{a}^\dagger_\mu$ and $\hat{a}^{}_\mu$
($\mu=1,2$) creates and annihilates an atom in the $\mu$th well,
respectively; and $\hat{n}^{}_\mu = \hat{a}^\dagger_\mu \hat{a}^{}_\mu$ is
the particle number operator. Here the parameter $\gamma$ denotes
the energy bias between the two wells, $T$ is the inter-well tunneling
strength, and $U$ is the interaction strength between atoms. The
Hamiltonian (\ref{eq:hamquan}) can describe not only the BEC in
double wells but also that in two hyperfine
states~\cite{Matthews,Li}. This model was conventionally studied in
a mean-field approach by replacing the expectation values of
annihilators in two different wells with two complex numbers $a_1$
and $a_2$, respectively~\cite{Biao,Lee}. In those works, the system
is essentially assumed to be of complete coherence, \ie, the
condensate stays in a completely coherent superposition state
$\ket{\psi_\textrm{coh}}=\frac{1}{N!}(a_1\hat{a}^\dagger_1+a_2\hat{a}^\dagger_2)^N
\ket{\textrm{vac}}$. Whereas, a realistic system of condensates may
be not always in complete coherence for various situations. It is
therefore worthwhile to study the dynamical properties of partially
coherent BECs in double wells.

As we know, it is convenient to introduce the single-particle
density matrix $\tilde{\rho}$ with entities
$\tilde{\rho}^{}_{\mu\nu}(t)=\ave{\hat{a}^\dagger_\mu(t)\hat{a}^{}_\nu (t)}$
where the expectation value is taken for the initial state of the system.
Clearly $\rho_{11}$ and $\rho_{22}$ represent the
population in the first and in the second well, respectively.
With the help of Heisenberg equation of motion for operators, one
can derive the dynamical equations for the elements of the
aforementioned density matrix by the mean-field approach
in the semiclassical limit.
\begin{eqnarray}\label{eq:dyeq}
i\frac{d\rho^{}_{11}}{dt}&=&-i\frac{d\rho^{}_{22}}{dt}
= -T(\rho^{}_{12}-\rho^{}_{21}),
\nonumber\\
i\frac{d\rho^{}_{12}}{dt}&=&-\gamma\rho^{}_{12}
+T(\rho^{}_{22}-\rho^{}_{11})
-UN(\rho^{}_{11}-\rho^{}_{22})\rho^{}_{12},
   \nonumber\\
i\frac{d\rho^{}_{21}}{dt}&=&
\gamma\rho^{}_{21}-T(\rho^{}_{22}-\rho^{}_{11})
+UN(\rho^{}_{11}-\rho^{}_{22})\rho^{}_{21},
\nonumber\\
\end{eqnarray}
where $\hbar$ is set to unit and
$\rho^{}_{\mu\nu}=\tilde{\rho}_{\mu\nu}/N$. The conservation of
particle number requires that $\rho^{}_{11}+\rho^{}_{22}=1$.

As the $2\times 2$ density matrix can be expanded as $\rho=(I + \vec
P\cdot\vec\sigma )/2$ with $\sigma$'s the Pauli matrices and $\vec
P$ a vector inside the so-called Bloch sphere. Since $|\vec P|=1$
refers to a pure state (completely coherent superposition) while
$|\vec P|<1$ refers to a mixed state (partial coherence), one can
measure the degree of coherence by $|\vec P|^2\equiv \eta$. Thus a
natural definition of the degree of coherence is given
by~\cite{Anthony}
\begin{equation}\label{eq:coherence}
\eta=2\textrm{Tr}\rho^2-1,
\end{equation}
which is an important quantity that affects the dynamical features,
such as Rabi oscillation, self-trapping, etc. With the help of the
mean-field dynamical Eqs. (\ref{eq:dyeq}) for the
single-particle density matrix, the dynamics of the system can be
investigated. We know that the ratio of the interaction strength $U$
to the tunneling strength $T$ determines three distinct regimes, namely,
the Rabi regime, $UN/T\ll 1$, the Josephson regime, $1\ll UN/T\ll
N^2$, and the Fock regime, $N^2 \ll UN/T$. The system manifests
different dynamical features in different regimes, which will be
given in the following sections.

\section{Evolution of the degree of coherence in the Fock regime}
\label{sec:coherence}

To study the evolution of the degree of coherence,
we can conveniently introduce the pseudospin operators,
\begin{eqnarray}
\hat{J}_z=\frac{1}{2}(\hat{a}^\dagger_1\hat{a}^{}_1-\hat{a}^\dagger_2
\hat{a}^{}_2),\nonumber\\
\hat{J}_x=\frac{1}{2}(\hat{a}^\dagger_1\hat{a}^{}_2+
\hat{a}^\dagger_2\hat{a}^{}_1),\nonumber\\
\hat{J}_y=-\frac{i}{2}(\hat{a}^\dagger_1\hat{a}^{}_2-
\hat{a}^\dagger_2\hat{a}^{}_1),\nonumber
\end{eqnarray}
which obey commutation relations for the angular momenta
$[\hat{J}_j,~
\hat{J}^{}_k]= i\epsilon_{_{jkl}}\hat{J}_l$,
and fulfil
\begin{equation*}
\mathbf{\hat{J}}^2=\hat{J}_x^2+\hat{J}_y^2+\hat{J}_z^2
=\frac{N}{2}\Bigl(\frac{N}{2}+1\Bigr),
\end{equation*}
for systems of $N$ bosons. In terms of these pseudospin operators,
the Hamiltonian (\ref{eq:hamquan}) can be written as,
\begin{equation}\label{eq:hamimom}
\hat{H}=\gamma\hat{J}_z+U\hat{J}_z^2-2T\hat{J}_x.
\end{equation}
This implies that the dynamical properties are determined only
by the direction of the pseudospin $\mathbf{J}$ since its
magnitude is fixed on $N/2$.

We know that the eigenvalue of $\hat{J}_z$ is $(n_1-n_2)/2$, which
refers to the population imbalance between the two wells. For
convenience, let us introduce the canonically conjugated operator
$\hat{\varphi}$ of $\hat{J}_z$ to characterize the relative phase,
which can be regarded as the angle of $\hat{\mathbf{J}}$ in the
$x$-$y$ plane in the angular-momentum picture. According to
Ref.~\cite{Anthony}, the operator $\hat{\varphi}$ can be defined
through $\hat{E}\equiv \exp{i\hat{\varphi}}$, where
\begin{equation}
\hat{E}=[(N/2-\hat{J}_z)(N/2+\hat{J}_z+1)]^{-1/2}(\hat{J}_x+i\hat{J}_y).
\end{equation}
Such a definition satisfies $[\hat{J}_z,
\hat{E}]=\exp{i\hat{\varphi}}$ which is consistent with  the
condition $[\hat{J}_z, \hat{\varphi}]=-i$, so that $\hat{J}_z$ and
$\hat{\varphi}$ are canonically conjugated to each other. Then, in
terms of $\hat{J}_z$ and $\hat{\varphi}$,
equation~(\ref{eq:hamimom}) can be rewritten as
\begin{equation}\label{eq:HT}
\hat{H}=U\hat{J}^2_z+\gamma\hat{J}_z-
TN\sqrt{1-\frac{4\hat{J}_z}{N^2}}\cos{\hat{\varphi}}.
\end{equation}

In the Fock regime, the tunneling strength is much smaller than the
interaction one between atoms, \ie, $U/(NT)\gg 1$, hence the last
term in Eq.~(\ref{eq:HT}) can be neglected. Such a condition can be
satisfied in experiment through increasing the distance between the
two wells or enhancing the height of the potential barrier
separating the two wells. Then the dynamical equations in this
regime become,
\begin{eqnarray}
\frac{d\hat{J}_z}{dt}=0,\quad\quad
\frac{d\hat{\varphi}}{dt}=2U\hat{J}_z+\gamma,\nonumber
\end{eqnarray}
which means that the difference in particle numbers between the two
wells is fixed  but the relative phase
$\hat{\varphi}=2U\hat{J}_zt+\gamma t$ evolves with time. According
to Eq.~(\ref{eq:coherence}) and the definition of pseudospin
operators, the degree of coherence can be written as
\begin{equation*}
\eta=\frac{4}{N^2}(\ave{\hat{J}_z}^2+|\ave{\hat{a}^\dagger_1\hat{a}_2}|^2).
\end{equation*}
From the definition of $\hat{\varphi}$, one can also get
\begin{equation*}
\ave{\hat{a}^\dagger_1\hat{a}_2}=
  \frac{N}{2}\ave{(1-\frac{4}{N^2}\hat{J}_z^2)^{1/2}\exp{i\hat{\varphi}}}.
\end{equation*}
Since the expectation value of $\hat{J}_z$ is fixed in the extreme
Fock limit $T=0$, the evolution of the degree of coherence only
depends on that of $\hat{\varphi}$. For example, if the particle
numbers in the two wells are the same at the initial time, the
expectation value of $\hat{J}_z$ will be always fixed on zero in the
extreme Fock limit. In this case, the degree of coherence becomes
$\eta=4|\ave{\hat{a}_1^\dagger\hat{a}_2}|^2/N^2\approx|
\ave{\exp{i\hat{\varphi}}}|^2$,
which reflects that the evolution of the degree of coherence is
determined by the interaction strength $U$, the detuning $\gamma$
and the initial state of the system.

As we know, the Fock bases $\ket{l}$ composing the
$(N+1)$-dimensional Hilbert space of the  system are
 \begin{eqnarray}
\ket{l}=\ket{\frac{N}{2}+l,\frac{N}{2}-l},\quad
 l=-\frac{N}{2},-\frac{N}{2}+1,\ldots,\frac{N}{2},
 \end{eqnarray}
where $l$ corresponds to the quantum number of
$\hat{J}_z$ denoting the half of the difference in the particle
numbers between the two wells. Since any initial state can be
written as
\[\ket{\Psi(0)}
=\sum_{l=-N/2}^{l=N/2}\psi_l\ket{l},
\]
the degree of coherence can be obtained as long as $\psi_l$ is given.
Taking
$\psi_l=e^{-(l-\delta)^2/l_0 ^2}\big/(\pi l_0 ^2/2)^{1/4}$ as an
example and replacing the sum over $l$ by an integral,
one can obtain the degree of coherence.
If keeping the lowest order term in the Taylor expansion in the
calculation of the expectation value of
$(1-\frac{4}{N^2}\hat{J}_z^2)^{1/2}\exp i\hat\varphi$, we have
\begin{equation}\label{eq:excohere}
\eta(t)\simeq\frac{4}{N^2}\delta^2+\exp{(-l_0 ^2U^2t^2)}.
\end{equation}
This expression is valid when the width of the
Gaussian distribution is much smaller than $N$, \ie, $l_0\ll N$.
Clearly, the degree of coherence decays with  time as long as there
is interaction between atoms. Such result tells us that one can
prepare systems with different degrees of coherence through changing
the evolution time $t$ in the extreme Fock regime. From
Eq.~(\ref{eq:excohere}), we can also find that the larger the $l_0 $
is, the faster the degree of coherence will decay. Then one can
suppress the decay through decreasing the value of $l_0 $.
Note that in the above calculation of $\eta(t)$,
we used $\hat{\varphi}=2U\hat{J}_zt+\gamma t$ which is valid
only for the extreme Fock regime. In the other regimes, one needs to solve Eq.~(\ref{eq:dyeq})
either analytically or numerically without any assumption.

\section{The influence of degree of coherence
on the Rabi oscillation}\label{sec:rabi}

Rabi oscillation is an important phenomenon reflecting the coherent
property of a system, which has been discussed
recently~\cite{Syassen,Cunha,Jai}. In Ref.~\cite{Jai} the initial
state was assumed to be a pure state, \ie, $\rho_{11}=1$, implying
$\eta=1$. Whereas, according to our formulation in
Sec.~\ref{sec:coherence}, one can prepare states with different
degrees of coherence, which makes Rabi oscillation worthwhile to be
studied from a new angle of view. Now we study the dynamics of Rabi
oscillation in terms of  the density matrix and give the relation
between the amplitude of the Rabi oscillation and the degree of
coherence for BECs in double wells.

\begin{figure}[h]
\includegraphics[width=90mm]{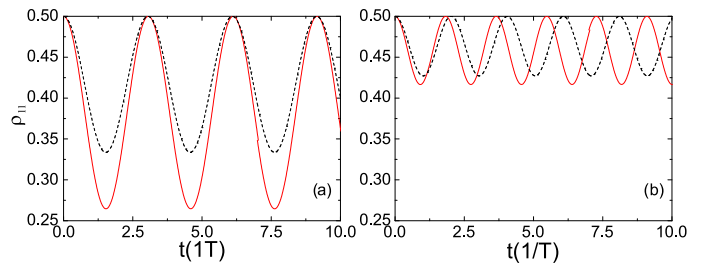}
\caption{\label{fig:rabiosc}(Color online) The Rabi oscillation of
the particle distribution with different degrees of coherence for
(a) $U/T=0$ and for (b) $UN/T=4$. The parameters are $
\gamma/T=0.5$ and $\eta=0.5$ (dot line), $1$ (solid line).}
\end{figure}

In the extreme Rabi limit, $U=0$, equations (\ref{eq:dyeq}) are
reduced to Bloch equations that can be solved analytically. For example, we consider a initial state described by the density matrix
$\rho_{11}^{}(0)=\rho_{22}^{}(0)=1/2$ and
$\rho_{12}^{}(0)=\rho_{21}^{}(0)=c$, we can obtain the solution of
Eqs.~(\ref{eq:dyeq}),
\begin{eqnarray}\label{eq:soR}
\rho_{11}^{}&=&\frac{1}{2}-\frac{2c\gamma
T}{\Omega^2}+\frac{2c\gamma
T}{\Omega^2}\cos{(\Omega t) },\nonumber\\
\rho_{22}^{}&=&\frac{1}{2}+\frac{2c\gamma
T}{\Omega^2}-\frac{2c\gamma
T}{\Omega^2}\cos{(\Omega t) },\nonumber\\
\rho_{12}^{}&=&c-\frac{c\gamma^2}{\Omega^2}
+\frac{c\gamma^2}{\Omega^2}\cos{(\Omega
t)}+i\frac{c\gamma}{\Omega}\sin{(\Omega t)},\nonumber\\
\rho_{21}^{}&=&c-\frac{c\gamma^2}{\Omega^2}
+\frac{c\gamma^2}{\Omega^2}\cos{(\Omega
t)}-i\frac{c\gamma}{\Omega}\sin{(\Omega t)},
\end{eqnarray}
where $\Omega=(4T^2+\gamma^2)^{1/2}$, and $c$ is a real number
related to the degree of coherence, \ie, $c=\sqrt{\eta}/2$. From
Fig.~\ref{fig:rabiosc}, we can see that the particle numbers in the
two wells both periodically oscillate with time. The oscillation
amplitude $c\gamma T/\Omega^2$ diminishes with the decrease in the
degree of coherence. Here the relative phase between the condensates
in two wells is taken as zero, \ie, $\rho_{12}(0)$ is real at the
initial time. If the degree of coherence is fixed but the relative
phase is not zero, saying $\rho_{12}(0)=c\exp(i\phi)$ and
$\rho_{21}(0)=c\exp(-i\phi)$, we can also solve Eqs.~(\ref{eq:dyeq})
analytically (the expression of $\rho_{ij}(t)$ is omitted for saving
space). One can find that the amplitude of Rabi oscillation is
affected by the initial values of the degree of coherence as well as
the relative phase.
For the sake of comparison, we
plot a numerical result of the Rabi oscillation
in the presence of interaction ($U\neq 0$) in Fig.~\ref{fig:rabiosc} (b).
Comparing the two panels in Fig.~\ref{fig:rabiosc},
we can see that the interaction between
atoms will make both the amplitude and the period of the Rabi
oscillation decrease for the same initial states.

The above discussions tell us that one can obtain the degree of
coherence  through measuring the amplitude of the Rabi oscillation
in experiment. For example, like in the experiment~\cite{Weiman},
prepare a BEC in one hyperfine state and transfer one half of atoms
into the other hyperfine state through a two-photon pulse; turn off
the pulse to allow the system to evolve freely without the
inter-state tunneling (\ie, $T=0$) until $t=\tau_0$. Then turn on a
pulse which makes the atoms tunnel between the two states and
measure the amplitude of the Rabi oscillation.
Because the
system is in the extreme Fock regime when $t<\tau_0$, based on the
calculation given in Sec.~\ref{sec:coherence}, we can have
$\rho_{11}^{}=\rho_{22}=1/2$ and
$\displaystyle\rho_{12}^{}=\frac{1}{2}\exp(i\gamma \tau_0)\exp(-l_0
^2U^2\tau_0^2/2)$  at the  time $\tau_0$.
Such a state can be actually prepared
as the initial state of the subsequent Rabi oscillation procedure.
Note that the expression of $\rho_{12}$ can be also
rewritten as $\rho_{12}=\sqrt{\eta}\exp(i\gamma\tau_0)/2$
considering the definition of $\eta$. Since the relative phase
$\gamma \tau_0$ is determined, the amplitude of  Rabi oscillation
only depends on the degree of coherence. The analytical result of
Eqs.~(\ref{eq:dyeq}) can not be obtained due to the existence of
interaction terms in the above example, so one can solve it
numerically. The degree of coherence can be determined through
fitting the experimental Rabi oscillation profiles with the
theoretical ones as the  degree of coherence is evolved.

\section{Self-trapping for the partially coherent system}\label{sec:selftrap}

As we know, the most prominent feature of atomic tunneling between
two wells is the nonlinear dynamics arising from the interaction of
atoms. Whereas, equations~(\ref{eq:dyeq}) can not be analytically
solved once the interaction terms are taken into account. In this
case, we solve Eqs.~(\ref{eq:dyeq}) numerically. In the numerical
calculation, we adopt a linearly time-dependent energy bias
$\gamma=\alpha t$ where $\alpha$ is a constant characterizing the
rate of the change of the energy bias $\gamma$. The initial values
are $\rho_{11}^{}=\rho_{22}=1/2$ and $\displaystyle\rho_{12}^{}
=\frac{1}{2}\sqrt{\eta}\exp{i\phi}$, where $\phi$ is the phase
difference between the condensates in the two wells at initial time.
Our results manifest that the initial relative phase and the degree
of coherence affect the dynamical properties of self-trapping.

\begin{figure}[h]
\includegraphics[width=66mm]{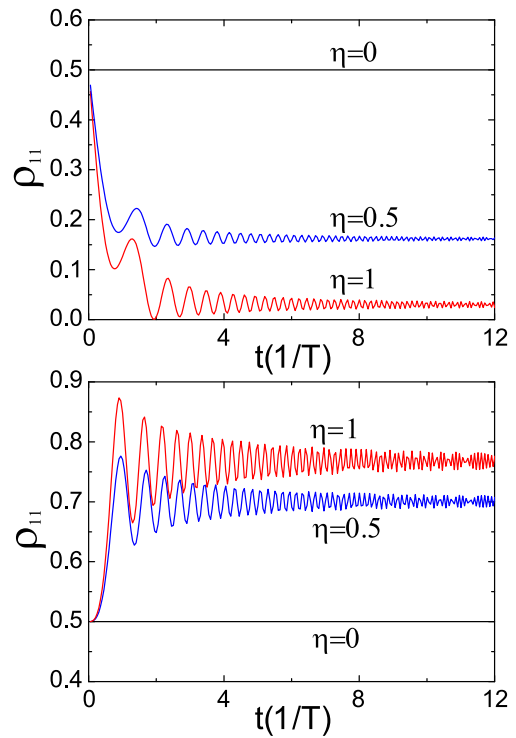}
\caption{\label{fig:population}(Color online) Time-evolution of the
population probability $\rho_{11}^{}$ for different degrees of
coherence and initial relative phases ($\phi=\pi/3$ for the top
panel and $\phi=\pi$ for the bottom  panel). The other parameters
are $UN/T=4$ and $\alpha/T^2=5$. }
\end{figure}
\begin{figure}
\includegraphics[width=70mm]{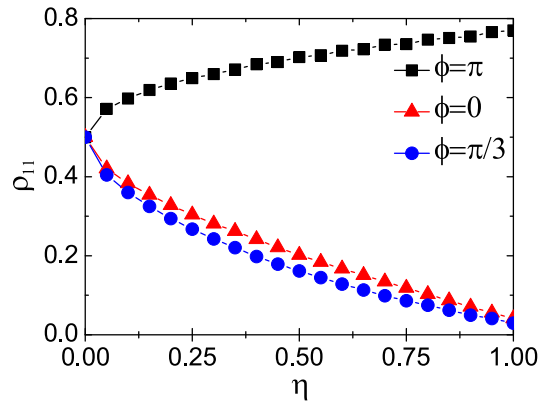}
\caption{\label{fig:eta}(Color online) The dependence of the final population
probability on the degree of coherence. The parameters are the same
as in Fig.~\ref{fig:population}. }
\end{figure}
\begin{figure}
\includegraphics[width=63mm]{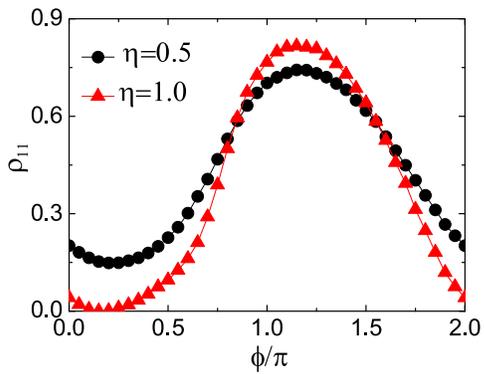}
\caption{\label{fig:phi} (Color online) The dependence of the final
population probability on the initial relative phase $\phi$. The
parameters are the same as in Fig.~\ref{fig:population}. }
\end{figure}

In Fig.~\ref{fig:population}, we plot the time-evolution of the
population $\rho_{11}^{}$ for different initial values of the degree
of coherence and relative phase. From this figure, we can see that
the system exhibits the phenomenon of self-trapping if the system is
coherent, \ie, $\eta\neq 0$, but the atoms do not tunnel between the
two wells for the incoherent system, \ie, $\eta=0$. We can also find
that most of the atoms favorite to stay in the right well (\ie,
$\rho_{22}>\rho_{11}$) at the final time for $\phi=\pi/3$, which is
contrast to the case for $\phi=\pi$. The difference between the two
panels of Fig.~\ref{fig:population} implies that the initial
relative phase can affect the phenomenon of self-trapping. It
depends on the energy bias and the initial relative phase that in
which well the atoms prefer to stay at the final time. When the
initial relative phase is fixed, figure~\ref{fig:population} shows
that the larger the degree of coherence is, the larger the
population difference between the two wells at the final time will
be. This is confirmed by Fig.~\ref{fig:eta} where the final
population probability versus  the degree of coherence is plotted
for three different values of $\phi$. In Fig.~\ref{fig:phi}, we plot
the relation between the final population probability and the
initial relative phase for different degrees of coherence. This
figure confirms that the initial relative phase can affect the final
distribution of atoms.

\begin{figure}[h]
\includegraphics[width=82mm]{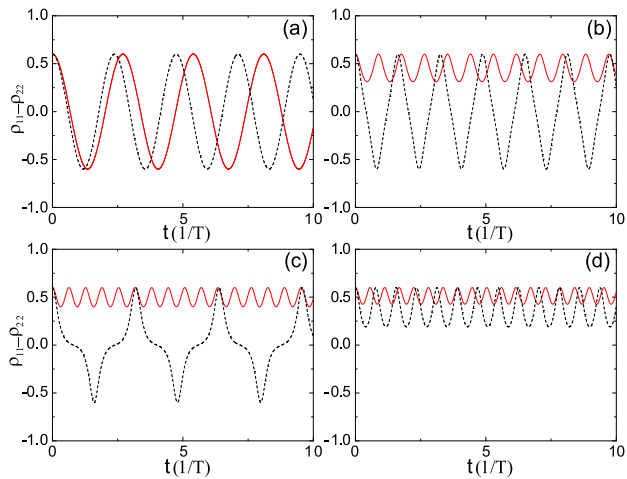}
\caption{\label{fig:resu}(Color online) Time-evolution of
population imbalance for a symmetric double-well potential with different degrees of coherence.  The initial states are $\rho_{11}=0.8$, $\rho_{22}=0.2$, and $\rho_{12}=0.4$ (dot line), $\rho_{12}=0.2$ (solid line). The parameters are (a) $UN/2T=1$, (b) $UN/2T=8$,
(c) $UN/2T=9.99$, and (d) $UN/2T=11$.}
\end{figure}
The emergence of the self-tapping phenomena
is dependent on the initial state and the systems parameters,
\eg, the interaction strength $U$, the inter-well tunneling strength,
and the energy bias $\gamma$ between two wells.
Such a phenomenon was investigated in a symmetric double-well potential~\cite{smerzi},
\ie, $\gamma=0$,
and also in a double-well potential with a periodic modulation,
\ie, $\gamma\propto \sin{\omega t}$~\cite{Wang2}.
According to Ref.~\cite{smerzi},
we can find that the self-trapping phenomena
occur only when the interaction strength is larger
than the critical value
$U_c\propto [\sqrt{1-z(0)^2}\cos{\phi(0)}+1]T/[Nz(0)^2]$
in the case of
$\gamma=0$, where $z(0)$ refers to the initial population difference
$\rho_{11}(0)-\rho_{22}(0)$ and $\phi(0)$ the initial relative phase
between the condensates in the two wells.
Therefore, for the initial states  we considered afore,
if there is no  energy bias,
the system can not exhibit the self-trapping phenomena due to the
critical interaction strength $U_c\rightarrow\infty$ for $z(0)=0$, $\phi(0)\neq \pi$\cite{foot}. For the initial states with population
imbalance (\ie, $z(0)\neq 0$), we plot the numerical solutions of Eqs.~(\ref{eq:dyeq}) for a symmetric double-well potential with different degrees of coherence in Fig.~\ref{fig:resu}.
Such a figure shows that the degree of coherence  affects the critical interaction
strength $U_c$ above which the system can exhibit the self-trapping phenomena in a symmetric double-well potential.
The system investigated in Ref.~\cite{smerzi} corresponds to the case when the degree of coherence $\eta=1$
in our discussion (see the dot line in Fig.~\ref{fig:resu}).

\section{Summary and Discussion}\label{sec:sum}

In the above, we considered Bose-Einstein condensates in double
wells with different degrees of coherence. With the help of
single-particle density matrix, we studied the dynamical properties
of partially coherent Bose-Einstein condensates in double wells and
showed that the degree of coherence is a useful parameter that
affects the dynamical features. We investigated the system in
different regimes and found that the degree of coherence can affect
the dynamical properties of the system significantly. In the Fock
regime, we mainly studied the time evolution of the degree of
coherence by introducing the pseudospin operators and showed that
the degree of coherence decays in exponential form of the square of
time. In the Rabi regime, we studied the effects of the initial
relative phase $\phi$ and degree of coherence $\eta$ on Rabi
oscillation and showed that the amplitude of Rabi oscillation is in
proportional to the square root of $\eta$ in the case of $\phi=0$.
According to the relevant result, the degree of coherence is
expected to be determined  through measuring the amplitude of Rabi
oscillation. Because the existence of nonlinear terms of interaction
makes the dynamical equations in the Josephson regime not solvable
analytically, we solved those equations numerically and found  that
the self-trapping phenomenon also exists for partially coherent BEC
systems in double wells. A more fascinating feature is that which
well the particles will stay in at the final time largely depends on
the initial relative phase. This result is different from that in
the previous works~\cite{Wang}, where the particles are in the same
well at the initial time such that the initial relative phase
between BECs in two wells does not affect the dynamical property of
systems explicitly.

The work is supported by NSFC under Grant No. 10674117 and No. 10874149,
and partially by PCSIRT under Grant No. IRT0754. The authors
acknowledge interesting discussions with L. B. Fu and J. Liu.


\begin{thebibliography}{99}
\bibitem{foot}
For the initial state of $z(0)=0$ and $\phi(0)=\pi$, if there is no the energy bias,
the system will always stay at the initial state which implies that the system can not exhibit the self-trapping
phenomena.
\end{thebibliography}
\end{document}